\documentstyle[prl, aps, twocolumn, epsf, psfig]{revtex}
\begin{document}
\draft
\twocolumn[\hsize\textwidth\columnwidth\hsize\csname@twocolumnfalse\endcsname
\title{Electrical Conductivity of Lithium at Mbar Pressures}
\author{Marina Bastea\cite{mb} and Sorin Bastea}
\address{Lawrence Livermore National Laboratory, P. O. Box 808, 
Livermore, CA 94550}
\maketitle
\begin{abstract}
We report measurements of the electrical conductivity of a liquid alkali metal - lithium - 
at pressures up to 1.8 Mbar and 4-fold compression, achieved through shock compression experiments. 
We find that the results are consistent with a departure of the electronic properties 
of lithium from the nearly free electron approximation at high pressures.
\end{abstract}
\pacs{PACS numbers: 72.15.-v, 62.50.+p, 71.22.+i}
]
At one atmosphere lithium is the archetype of a ``simple'' metal, i.e. one in which the electronic 
valence states are well separated energetically from the tightly bound core states. 
It has been known for a long time that its electronic properties at these ambient conditions 
are well described within the nearly free electron picture \cite{seitz}, which is a 
cornerstone of the theory of simple metals. While it has been generally assumed that increased 
pressure would only improve the accuracy of this description, in recent years theoretical 
calculations suggested that the opposite may be true \cite{boettger}. This has culminated 
very recently with the striking prediction that lithium, long viewed as the simplest of all 
metals, exhibits at Mbar pressures a paired ground state 
with a semiconducting or insulating character \cite{neaton1}. The ensuing experimental work  
has provided results on the decrease of optical reflectivity upon compression  
\cite{mori,hemley1}, on the increase of the electrical resistivity \cite{fortov} with 
pressure up to 0.6 Mbar, and on the existence of symmetry-breaking structural transitions around 0.5 Mbar 
\cite{hanfland}, all in broad agreement with this prediction. It is perhaps worth noting 
that even before the nearly free electron model was proposed, Bridgman's high pressure 
experimental work showed that the resistivity of lithium increases with pressure, both 
in the solid and in the liquid \cite{bridgman}. 
Here we report measurements of the electrical resistivity of lithium 
at pressures up to 1.8 Mbar, achieved through shock compression experiments. 
The results are consistent with a departure of the electronic properties 
of lithium from the nearly free electron approximation at high pressures, and with ionic 
pairing correlations in the Mbar regime ($1 Mbar = 100 GPa$). 
  
We measured the electrical resistance of high purity $99.995\%$ lithium samples 
quasi-isentropically compressed starting from the solid with density $d_0=.531 g/cm^3$ 
at room temperature and $P_0=1 bar$. The quasi-isentropic compression was achieved 
through multiple reflections of a shock wave \cite{oxygen} between two sapphire 
single-crystals which encapsulate the lithium sample. 
The initial shock wave was generated by the impact of Al or Cu projectiles moving at 
$3$ to $7 km/s$ onto the experimental cell (target) containing the samples. The targets 
were specifically designed to create and maintain steady state conditions at the final 
pressure for time durations exceeding $100ns$, during which the measurements 
were taken.

Due to the extreme reactivity of lithium, the experimental cells were assembled 
inside an Ar atmosphere glove-box, and designed to be hermetically sealed after assembly. 
Pre-shock sample resistances were within $5\%$ of the ideal room temperature values and were monitored 
continuously after the targets were removed from the controlled environment to ensure that 
the sample quality did not degrade. In order to increase the electrical resistance and better 
control current flow, we used parallelipipedic samples with dimensions 
$length\times width = 20mm\times 3mm$ and thicknesses varying between $.2mm$ and $.3 mm$. 
The $10:1$ and bigger aspect ratios used insured that lateral rarefaction effects were minimal.  
\begin{figure}
\centerline{\psfig{file=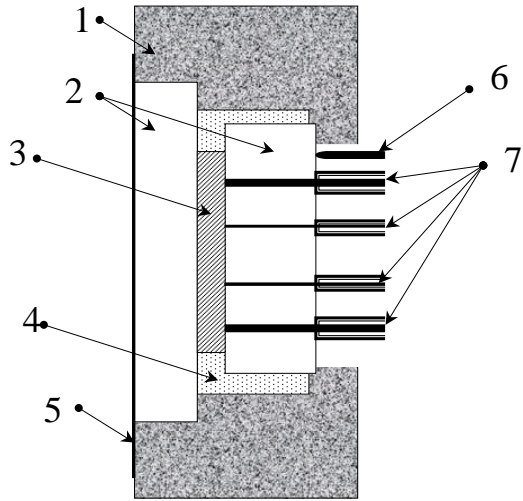,width=2.75truein,angle=270}}
\caption{Schematic representation of the experimental cell (not to scale). (1) Al target body; (2) 
single-crystal sapphire shock anvils; (3) Lithium sample; (4) Polyethilene filler - provides the 
electrical insulation around the sample; (5) Al or Cu foil matching the projectile material to 
provide screening of electrical discharges produced upon impact between the metallic flyer and the 
target; (6) triggering pins; (7) 4 Cu electrodes for the conductivity measurements.}
\label{target}
\end{figure}
The electrical resistance of the lithium samples was measured using a four probe technique, which 
virtually eliminates the need for the contact resistance correction. 
The electrodes were gold-plated, oxygen-free 
Cu wires, inserted through the back sapphire anvil, see Fig.\ref{target}. The diameter and 
positioning of the outer electrodes were optimized in order to provide uniform current injection 
into the lithium sample, and to minimize the contact resistance and distortions of the pressure 
profile in the measurement region. The voltage probes - inner electrodes - were very thin and 
placed $3-6mm$ apart as needed to maximize the sample resistance and the accuracy of the voltage 
measurements. Special attention was given to the soldering process and shielding of the electrodes 
in order to improve the signal-to-noise ratios. Triggering of the data acquisition system was 
provided by time-of-arrival shock sensors placed outside the sample space in order to eliminate 
any interference with the measurements. The details of the electronic circuitry used in 
these experiments are similar with the ones described in \cite{art-cond}.
\begin{figure}
\centerline{\psfig{file=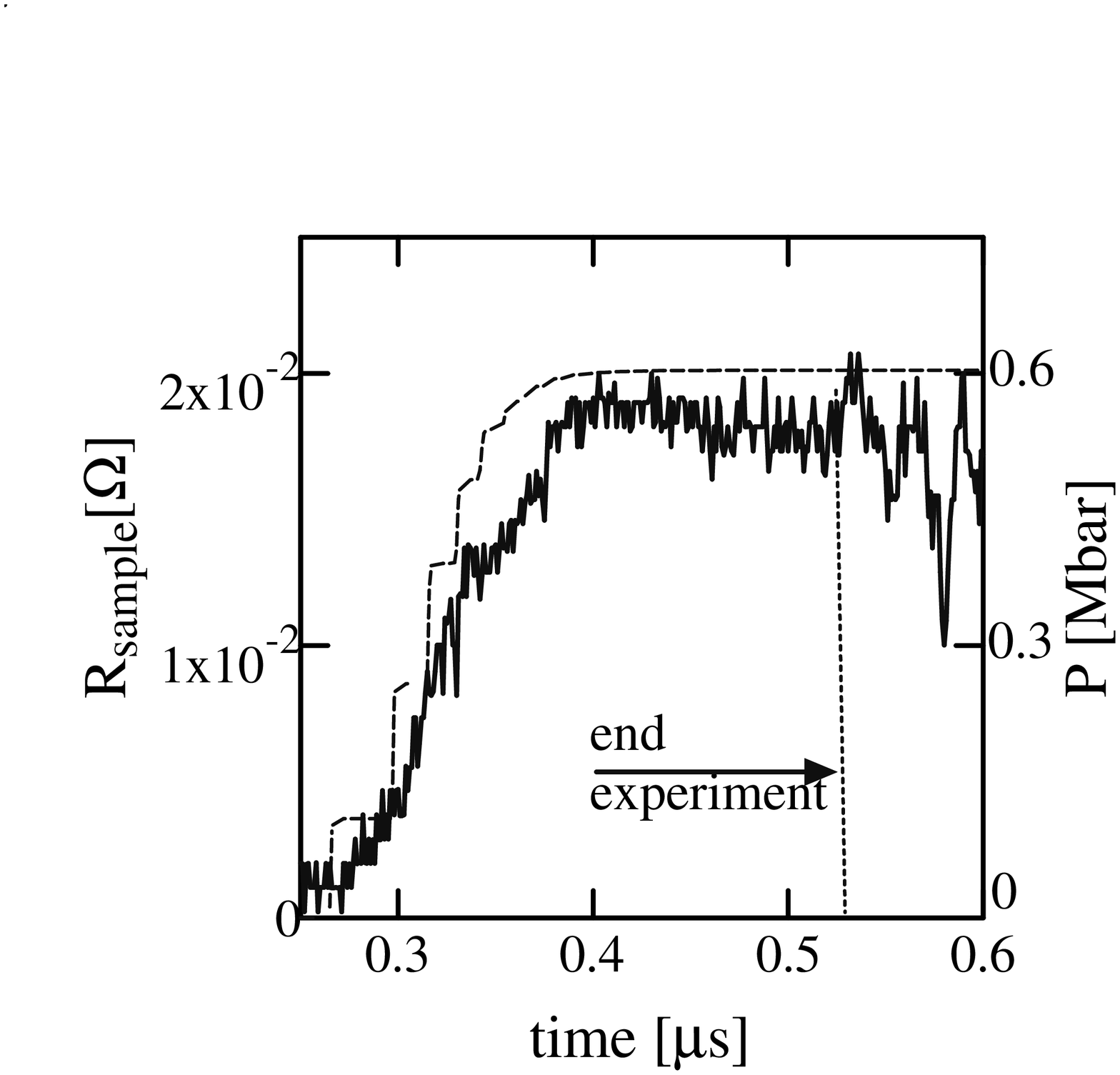,width=3.0truein}}
\caption{Example of a typical experimental trace. Measured sample resistance (scale on left).vs.time 
- continuous line; calculated pressure (scale on right).vs.time - dashed line.}
\label{sample-signal}
\end{figure}
The sample pressures at final, steady-state conditions were determined with $1\%$ accuracy from the 
measured projectile velocity using the shock impedance matching technique \cite{artSIM}. The 
time dependence of the pressure during compression was calculated using a 
one-dimensional hydrodynamic code in which the projectile and shock-anvils were modeled by 
Mie-Gruneisen equations of state (EOS) \cite{Marsh}. The EOS for the lithium sample 
was obtained by least-squares fits of a ratio of polynomials to tabular data based on extensive 
shockwave and static compression experimental results \cite{DY-eos}. We note that all the conductivity 
results reported in this paper are in the liquid region of the phase diagram. The densities 
and temperatures attained in the experiments range approximately from 2 to 4 times the normal density, 
and from $2000K$ to $7000K$.

In the experiment we monitored the voltage drop across the lithium sample as the pressure 
increased and a known constant current was passed through, see Fig. \ref{sample-signal}. The 
accuracy in the current and voltage measurements was higher than $1\%$ and $5\%$, respectively. 
The lithium electrical resistivity was determined from the sample resistance using 
$R=\rho \times l/S$, where the distance $l$ between the voltage probes was measured to $0.1\%$ 
precision. The cross-section of the sample, $S=w\times th$, was determined by the known 
width $w$ and calculated thickness $th$, with a conservatively estimated precision of 
$10\%$ arising mainly from density uncertainties.
\begin{figure}
\centerline{\psfig{file=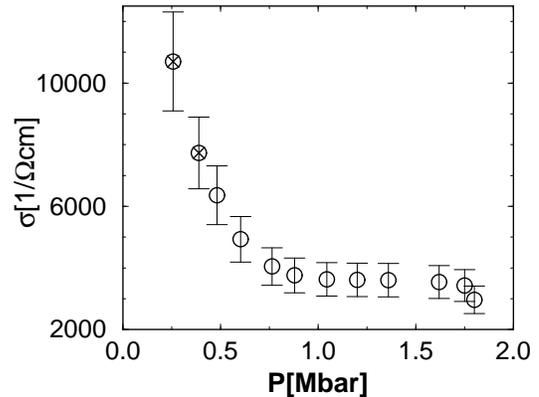,width=2.75truein}}
\caption{Measured electrical conductivity of lithium as a function of pressure. 
Open symbols - data extracted from the final, steady-state conditions; crossed 
circles - data extracted from intermediate states. Note: $1 Mbar = 100 GPa$.}
\label{sigmap}
\end{figure}
The measured electrical resistivity exhibits three main regimes as a function of pressure, 
see Fig. \ref{sigmap}. Up to approximately $1 Mbar$ and $3-fold$ compression the electrical conductivity 
of lithium decreases steeply from its $10^5(\Omega cm)^{-1}$ value at normal pressure to about
$3.6\times 10^3(\Omega cm)^{-1}$ at $1 Mbar$, which is consistent with increased scattering of 
the charge carriers in the compressed material. This regime is followed by a much slower 
variation of the conductivity with pressure, which extends to $\simeq 1.6 Mbar$ and almost 
$4-fold$ compression. At even higher pressures, although the error bars overlap, the electrical 
conductivity appears to depart from the previous slow varying behavior. 

In order to interpret the results we turn to the general framework provided by the 
electron-ion pseudopotential formalism and the Ziman conductivity theory \cite{ziman}. 
The pseudopotential concept takes into account the near cancellation of the Coulombic 
interaction between the valence electrons and ions inside the core, due to orthogonality 
and exclusion effects. While generally non-local and energy dependent, in its most 
intuitive representation it is just a local potential \cite{ashcroft1}. Under the assumption 
of a weak pseudopotential the linear response theory applied to the uniform electron gas 
yields the total energy of the system and defines effective ion-ion and electron-ion 
interactions \cite{ashcroft2,shimoji}. 

The effective ion-ion pair potential is the sum of a direct Coulombic interaction,
and an indirect contribution due to the polarization of the electrons,
\begin{equation}
\Phi(r) = \frac{Z^2e^2}{r} + v_{ind}(r)
\end{equation}
\begin{equation}
v_{ind}(k) = \chi(k)|w(k)|^2
\end{equation}
\begin{equation}
\chi(k) = \frac{\chi_0(k)}{1 - \frac{4\pi e^2}{k^2}[1 - G(k)]\chi_0(k)}
\end{equation}
and the screened electron-ion potential is, 
\begin{equation}
v(k) = \frac{w(k)}{\epsilon(k)}
\end{equation}
\begin{equation}
\epsilon(k) = 1 - \frac{4\pi e^2}{k^2}[1 - G(k)]\chi_0(k)
\end{equation}
where $Z$ is the number of valence electrons per atom, $\chi(k)$ is the static response function, 
$w(k)$ is the bare electron-ion pseudopotential, $\chi_0(k)$ is the Lindhard polarizability, 
$\epsilon(k)$ is the static dielectric function and $G(k)$ is the local field factor of the 
electron gas, accounting for the exchange and correlation effects between the electrons 
\cite{shimoji}. 

The electrical resistivity follows from the assumption of a degenerate electron gas and weak 
electron-ion scattering, yielding the Ziman formula \cite{shimoji,ashcroft3,stevenson}:
\begin{equation}
\rho = \left(\frac{m^*}{m}\right)^2\frac{a_0\hbar}{e^2}\frac{4\pi^3Z}{a_0 k_F}{\int_0^1} v^2(y)S(y)y^3 dy
\label{Zimaneq}
\end{equation}
The deviations of the electronic density of states from the free electron values, due to the 
pseudopotential, are accounted for through the effective mass $m^*$ \cite{itoh,oosten,chan}; 
$S$ is the liquid-structure factor, $y = k/2k_F$, $k_F$ is the Fermi wavevector and $a_0$ is 
the Bohr radius.

As remarked in \cite{neaton1}  the strongly non-local character of the pseudopotential generates 
significant deviations of the electronic structure from the free electron form at high densities, 
reflected in part in the decrease of the occupied bandwidth, i.e. increase of the effective mass. 
We note that, under suitable assumptions, Eq. \ref{Zimaneq} allows us to estimate the effective 
mass $m^*$, by comparing with the measured experimental results. For the sake of simplicity and 
to gain additional insight (see below), we use 
the local empty core potential of Ashcroft \cite{ashcroft1}, $w(r)=0$ for $r<R_c$ and $-Ze^2/r$ 
above $R_c$. However, we do not expect that the use of a non-local potential for the 
calculation of the scattering integral will 
qualitatively alter the results \cite{oosten}. Values of $R_c$ typical for lithium range from 
$1.06a.u.$ \cite{ashcroft3} to $1.44a.u.$ \cite{canales}, depending on the local field correction employed. 
We use the homogeneous electron gas local field factor $G(k)$ determined by diffusion Monte 
Carlo simulations \cite{moroni}, and find that $R_c=1.26a.u.$ reproduces the experimental values 
of the conductivity of liquid lithium close to its triple point \cite{chi} with an effective mass 
only slightly bigger than unity \cite{oosten}. We also note that the normal 
pressure effective ion-ion potential so obtained compares favorably with other potentials that have 
been used to model liquid lithium \cite{canales}. The liquid-structure factor $S(k)$ of liquid 
metals has been usually assumed to be well reproduced by the hard-sphere model with some appropriate 
packing fraction \cite{ashcroft2}. Here we determine $S(k)$ numerically based on the effective 
ion-ion interaction $\Phi(r)$, using the perturbative hypernetted-chain equation (PHNC) \cite{kang1}. 
PHNC has been successfully applied to various model systems, including liquid metals \cite{kang2}, and 
should be particularly appropriate at high densities. We note here that the temperature dependence of the 
scattering integral is embedded in $S(k)$. We calculate for example that a $20\%$ increase in temperature yields 
a $4\%$ decrease in the scattering integral at 2-fold compression, and only about $2\%$ at 4-fold 
compression. This is due to the fact that at high densities the liquid structure is determined 
mainly by the ion-ion repulsions \cite{lekner}, whose role only increases with increased compression.    
 
\begin{figure}
\centerline{\psfig{file=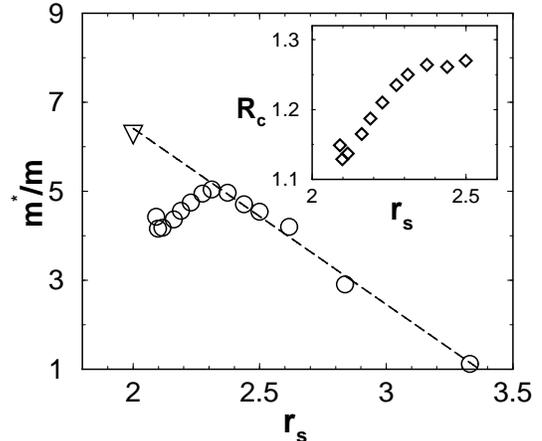,width=2.75truein}}
\caption{Electronic effective mass as a function of $r_s$ ($r_s = (3/4\pi n_e a_0^3)^{1/3}$, where $n_e$ is the valence 
electrons number density, with one electron/atom; $m$ is the electron mass). 
Open circles - estimated effective mass for $R_c=1.26a.u.$; dashed line - 
extrapolation of the lower pressures regime; triangle - effective mass reported in Ref. [3]. 
Inset shows the variation of the effective core radius with $r_s$ along the dashed line. The units for 
$R_c$ and $r_s$ are a.u..}
\label{effmass}
\end{figure}

The results of the effective mass calculations based on the Ziman theory with $R_c=1.26a.u.$ 
are summarized in Fig. 
\ref{effmass}. We note that the initial steep decrease in conductivity with pressure corresponds to an 
increase of $m^*$ to values similar with the ones found in \cite{neaton1}, as lithium becomes 
less free-electron like upon compression. In this pressure range the reduction in conductivity is 
driven mostly by increased scattering due to decreased interatomic spacing, although the core 
electronic states appear to already come into play as evidenced by changes 
in the effective mass. 
In the next regime the slow variation of the conductivity, 
see Fig. \ref{sigmap}, is accompanied by a decrease of the Ziman estimated effective mass, 
Fig. \ref{effmass}. It is very likely in fact that the actual effective mass \cite{chan} increases 
monotonically under compression past the predicted pairing instability \cite{neaton1}. 
This is the behavior observed in LDA calculations of a 
dense lithium monolayer, which is believed to behave similarly with bulk lithium \cite{bergara}. 
We note therefore that the slow variation of the conductivity, along with the decrease of the 
Ziman estimated effective mass, point to a shrinking ionic core, which balances 
the increase in scattering due to increased density. In the final regime, as the 
conductivity appears to drop, the effective mass increases again. In the framework of an empty core 
model this suggests that the exclusionary effects of the ions on the scattering of the valence 
electrons become dominant over the density driven core decrease. To make these observations more 
quantitative, we extrapolate the lower density effective mass to the high density value reported 
in \cite{neaton1}, and determine the core radius $R_c$ that reproduces this trend - see inset 
to Fig. \ref{effmass}.

As discussed in \cite{neaton1,neaton2} in connection to several alkali metals, 
the importance of the core electronic states increases at high pressures, and along with it 
the magnitude of the pseudopotential. As a result, in the solid, symmetry-breaking distortions 
leading to ionic pairing may become energetically favorable \cite{neaton1}. In the liquid, 
a rising pseudopotential, i.e. a decreasing effective ionic core, see Fig. \ref{effmass},
should also mediate a strengthening of pairing correlations.
Due to the strong exclusionary effects of the ions on the valence electrons at high 
densities, such correlations maximize in turn the spaces available 
for the electrons in the interstitial regions, away from the areas of strong core 
overlap in-between the ions, with beneficial effects on the energy. 
This exclusion of the valence electrons by the ions seems to be consistent with the behavior 
of the conductivity at high compressions. Given the expected small effect of the 
temperature on the conductivity at high densities, the apparent conductivity drop and the 
behavior of the ionic core at the highest pressures could be interpreted 
as a decrease of the overall volume available for the electrons.

It may be interesting to see if the exclusionary effects mentioned above, that lead to a very 
non-uniform distribution of the valence electrons, translate into the analog of classical 
depletion forces \cite{vlachy}, enhancing the ionic pairing correlations in a mixture, e.g. LiH. 
Higher temperatures would ultimately destroy pairing correlations 
in the liquid. Estimates of the temperatures required should be possible, but need to rely 
on more detailed calculations, that could be tested against the experimental results 
presented here.

We thank N. Ashcroft, J. Neaton, G. Galli and R. Cauble for useful discussions, 
and D. Young for the lithium EOS table. M.B. warmly thanks A. Mitchell for sharing his 
life-long experience on shockwave experiments. We gratefully acknowledge the LDRD office for 
financial support. We also thank S. Caldwell, W.P. Hall, N.Hinsey, K. Stickle, L. Raper 
and T. Uphaus for assistance at the gas-gun facility, and L. Walkley for help with the set-up 
of the glove-box. This work was performed under the auspices of the U. S. Department of 
Energy by University of California Lawrence Livermore National Laboratory under Contract 
No. W-7405-Eng-48.

\end{document}